\documentclass[aip,apl,12pt,preprint]{revtex4-1}

\usepackage{amssymb}
\usepackage{amsmath}
\usepackage{graphicx}
\usepackage{dcolumn}
\usepackage{bm}
\usepackage[usenames,dvipsnames,svgnames,table]{xcolor}

\begin{document}

\title{Stochastic model of dispersive multi-step polarization switching in ferroelectrics
due to spatial electric field distribution}
\author{R.~Khachaturyan}
\email[Electronic address: ]{rubenftf@gmail.com}
\affiliation{Institut f\"{u}r Materialwissenschaft, Technische Universit\"{a}t Darmstadt,
Otto-Berndt-Stra{\ss}e 3, D-64287 Darmstadt, Germany}
\author{J.~Schulthei\ss}
\affiliation{Institut f\"{u}r Materialwissenschaft, Technische Universit\"{a}t Darmstadt,
Alarich-Weiss-Stra\ss e 2, D-64287 Darmstadt, Germany}
\author{J.~Koruza}
\affiliation{Institut f\"{u}r Materialwissenschaft, Technische Universit\"{a}t Darmstadt,
Alarich-Weiss-Stra\ss e 2, D-64287 Darmstadt, Germany}
\author{Y.A.~Genenko}
\email[Electronic address: ]{genenko@mm.tu-darmstadt.de}
\affiliation{Institut f\"{u}r Materialwissenschaft, Technische Universit\"{a}t Darmstadt,
Otto-Berndt-Stra{\ss}e 3, D-64287 Darmstadt, Germany}
\date{\today }

\begin{abstract}
A stochastic model for polarization switching in tetragonal ferroelectric ceramics is introduced,
which includes sequential 90$^{\circ}$- and parallel 180$^{\circ}$-switching processes and accounts 
for the dispersion of characteristic switching times due to a nonuniform spatial distribution of 
the applied field. It presents merging of the 
recent multistep stochastic mechanism (MSM) with the earlier nucleation limited switching (NLS)
and inhomogeneous field mechanism (IFM) models. The new model provides a much better description of 
simultaneous polarization and strain responses over a wide time window and a deeper insight into the
microscopic switching mechanisms, as is exemplarily shown by comparison with measurements on lead 
zirconate titanate.

\end{abstract}

\maketitle

Multi-step non-180$^{\circ}$ polarization switching events were experimentally observed in ferroelectrics 
since the 90's by diffraction techniques\cite{Li1994,Daniels2007,Daniels2014}, ultrasonic 
methods\cite{Yin2001}, and microscopy\cite{Hsieh2009,Xu2014}. Although the account of these processes is 
ultimately necessary to describe the electromechanical response of ferroelectrics to an applied electric 
field, they are not included in common statistical models of polarization response, such as the 
Kolmogorov-Avrami-Ishibashi (KAI)\cite{Kolmogoroff,Avrami,KAI,KAI2}, the nucleation limited switching 
(NLS)\cite{NLS,JoPRL2007,JoAPL2008} and the inhomogeneous field mechanism (IFM)\cite{IFM1,IFM2,Lee2016} 
models, dealing with statistically independent parallel 180$^{\circ}$ switching processes only.   
Furthermore, experimental results revealed that both 180$^{\circ}$ and non-180$^{\circ}$ switching 
events are required in order to describe the electrical and mechanical response of polycrystalline 
ceramics to an applied electric field pulse during polarization 
reversal\cite{Yamada1996,Ogawa1998,SchultheissAM2018}.
Understanding of switching mechanisms and, particularly, knowing the fractions of 180$^{\circ}$
and non-180$^{\circ}$ contributions, is necessary for optimization of piezoelectric properties
of materials\cite{Pramanick2011,Glaum2012}. To this end, the fraction of non-180$^{\circ}$ 
switching events in a tetragonal BaTiO$_3$ at room temperature was evaluated to be around 20\% by means 
of {\it in situ} X-ray diffraction\cite{Fancher2017}. Using the recently advanced multistep stochastic 
mechanism (MSM) model\cite{GenenkoPRB2018-2} for the analysis of simultaneous polarization and strain 
measurements, this fraction was found to be about 34\% in a tetragonal lead zirconate titanate 
(Pb$_{0.985}$La$_{0.01}$(Zr$_{0.475}$Ti$_{0.525}$)O$_3$) at room 
temperature.

A common difficulty for both experimental and theoretical statistical analysis consists in the 
possibility of hypothetical coherent non-180$^{\circ}$ processes, suggested by Arlt\cite{Arlt1997},
which do not contribute to macroscopic  strain and thus appear to be mechanically identical to the 
180$^{\circ}$ reversal events. Another origin of uncertainty of interpretation of experimental results 
within the MSM model consists in the simplifying assumption of the uniform electric field all over 
the system. This assumption does not allow explanation of dispersive polarization and strain responses 
at later switching stages\cite{GenenkoPRB2018-2}, which may result from the distribution of local 
switching times due to the spatially inhomogeneous distribution of the applied 
field\cite{IFM1,IFM2,Lupascu2004}. To account for this circumstance and to improve the description of 
the experiment\cite{GenenkoPRB2018-2}, in the present study the NLS and the IFM models are merged with 
the MSM model. This means that both sequential 90$^{\circ}$-and parallel 180$^{\circ}$-switching processes 
are deemed to be driven by a nonuniformly distributed applied field. Similar to the previous KAI, NLS and 
IFM models, the actual hybrid model neglects electric and elastic interactions between the switching 
regions during polarization reversal.

\begin{figure}[b]
\includegraphics[width=8.5cm]{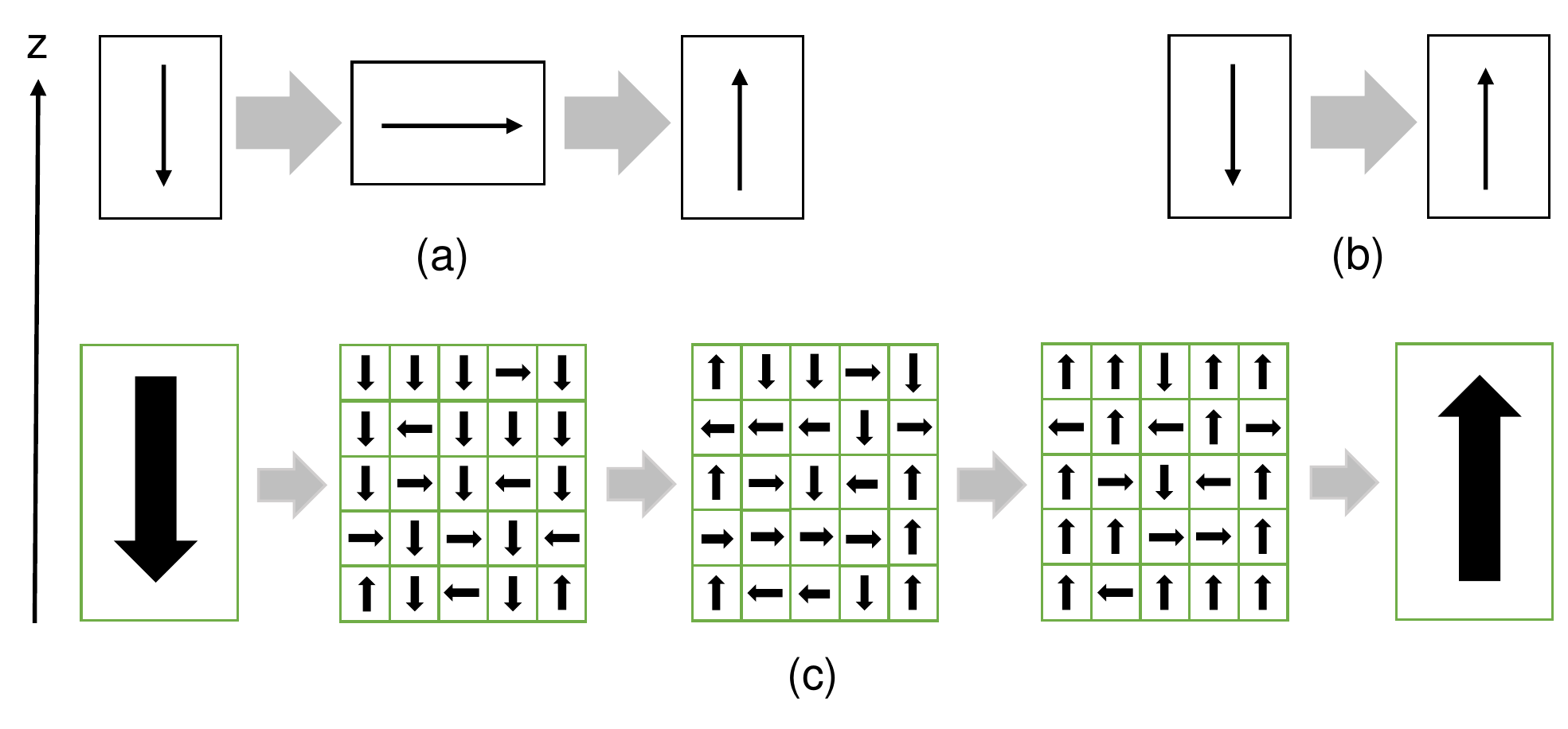}
\caption{ 
Changes in polarization and geometry of a unit cell due to idealized (a) squential 90$^{\circ}$ 
and (b) parallel 180$^{\circ}$ switching events. (c) Switching of a macroscopic sample by both
types of events. Green boxes represent independently-switching regions.}
\label{fig1}
\end{figure}
In this model, a poled polycrystalline tetragonal ferroelectric is assumed to be uniformly polarized 
in the negative $z$-direction, exhibiting a saturation polarization $-P_s$ (see Fig.~1(c)). When a 
reversed field is applied in the positive $z$-direction, the local polarization may experience two sequential 90$^{\circ}$ switching 
events (Fig.~1(a)) or a single 180$^{\circ}$ switching event (Fig.~1(b))\cite{SchultheissAM2018}.
The MSM model\cite{GenenkoPRB2018-2}
describes the macroscopic polarization response $\Delta p$ of this simplified system in $z$-direction  
by the formula 

\begin{align}
\label{dp-MSM}
\frac{\Delta p(t)}{2P_s} &= \eta \left\{1-\exp{\left[-\left(\frac{t}{\tau_1}\right)^{\alpha}\right]}\right\}
-\frac{\eta}{2} L_1(t)  \nonumber \\
&+ (1-\eta) \left\{1-\exp{\left[-\left(\frac{t}{\tau_3}\right)^{\gamma }\right]}\right\}
\end{align}
with a convolution of the 90$^{\circ}$ events
\begin{equation}
\label{L1-MSM}
L_1(t) =  \int\limits_0^t \frac{dt_1}{\tau_1}\alpha \left(\frac{t}{\tau_1}\right)^{\alpha -1}
\exp{\left[ -\left(\frac{t_1}{\tau_1}\right)^{\alpha} - \left(\frac{t-t_1}{\tau_2}\right)^{\beta} \right]}
\end{equation}
where the parameter $\eta<1$ quantifies the fraction of the 90$^{\circ}$-switching events, $\tau_i$ are the 
unique characteristic switching times for the first and the second 90$^{\circ}$-switching processes ($i=1$ 
and $2$, respectively) and the parallel 180$^{\circ}$- processes ($i=3$), while $\alpha, \beta$ and $\gamma$ 
are the Avrami indexes of the respective processes related to the dimensionality of the growing 
reversed polarization domains\cite{Avrami,KAI}. According to Merz\cite{Merz} and experimental observations 
on PZT ceramics\cite{JoPRL2007,JoAPL2008,IFM1,IFM2,SchultheissAM2018,GenenkoPRB2018-2} the field dependence 
of the characteristic switching times is adopted in the form
\begin{equation}
\label{Mertz}
\tau_i(E_a) = \tau _0 \exp{\left(E_A^{(i)}/E_a\right)} 
\end{equation} 
where $E_A^{(i)}$ are the activation fields for the above mentioned switching processes, and $E_a$ is the 
value of the applied uniform electric field. While in the framework of the MSM model regions of different 
length scales are assumed to switch statistically independent, this concept is able to account for the 
influence of microscopic parameters ({\it e.g.}, lattice inhomogenieties, defects, chemical dopants) 
directly reflected in the determined characteristic switching times and activation fields.

The time-dependent  change of the strain in $z$-direction is given in the MSM model by the formula 
\begin{equation}
\label{dS-MSM}
\Delta S_3(t) = \Delta S_{max}\eta L_1(t) +
2\varepsilon_0\varepsilon_{33}Q_{11} E_a ( \Delta p(t)-P_s )  
\end{equation}
with $L_1$ and $\Delta p$ functions of time defined by Eqs.~(\ref{L1-MSM}) and (\ref{dp-MSM}), respectively,
$\varepsilon_0$ and $\varepsilon_{33}$ the permittivity of vacuum and the relative pemittivity in 
$z$-direction, respectively, and $\Delta S_{max}$ the maximum negative strain. Voigt 
notations\cite{Newnham-book} are used for the components of the strain tensor ${\mathbf S}$ and the 
tensor of electrostriction ${\mathbf Q}$.

If now a nonuniform distribution of an applied electric field in a polycrystalline ceramic is taken into 
account, the local values $E$ of the electric field become also strayed around the value $E_a$ with some 
statistical distribution function $Z(E)$. Differently from the case of parallel switching 
processes\cite{IFM2,Lee2016},
the latter function cannot be easily derived from the polarization response in the case of sequential
processes. To keep the theory as simple as possible, the function $Z(E)$ may be assumed to have a Lorentzian
form, as adopted by Jo {\it et al.} in the NLS model\cite{JoPRL2007,JoAPL2008}. Furthermore, due to scaling 
properties of the polarization response well established in PZT and other 
ceramics\cite{IFM1,IFM2,ZhukovAPL2013,ZhukovJAP2014,ZhukovJAP2015,ZhukovJAP2016,ZhukovAPL2016}, 
it can be chosen in a scaling form with a dimensionless width of the field distribution $\kappa$:
\begin{equation}
\label{Lorentz}
Z(E) = \frac{1}{\pi E_a} \frac{\kappa }{(E/E_a-1)^2 +\kappa^2}. 
\end{equation}
Introducing Eq.~(\ref{Lorentz}) in Eq.~(\ref{dp-MSM}) an extended form results:
\begin{align}
\label{dp-MSM-IFM}
\frac{\Delta p(t)}{2P_s} =  \int\limits_0^{\infty}dE Z(E) 
\left\{
\eta\left[ 1-\exp{\left(-\left(\frac{t}{\tau_1(E)}\right)^{\alpha}\right)}\right]
\right.\nonumber\\
\left. +(1-\eta )\left[ 
1 -\exp{ \left( -\left(\frac{t}{\tau_3(E)}\right)^{\gamma }\right) } \right]
\right\} -\frac{\eta}{2}L_1(t)
\end{align}
with
\begin{align}
\label{L1-MSM-IFM}
&L_1(t) = 
\int\limits_0^t dt_1 \int\limits_0^{\infty}dE_1 Z(E_1)\frac{\alpha }{\tau_1(E_1)}
\left(\frac{t_1}{\tau_1(E_1)}\right)^{\alpha -1}\nonumber\\
&\times \int\limits_0^{\infty}dE_2 Z(E_2)
\exp{\left[
-\left(\frac{t}{\tau_1(E_1)}\right)^{\alpha}
- \left(\frac{t-t_1}{\tau_2(E_2)}\right)^{\beta}
\right]} 
\end{align}
where $E, E_1$ and $E_2$ present local field values.
For the strain response in this case the previous formula (\ref{dS-MSM}) still applies;
however, with the functions $\Delta p(t)$ and $L_1(t)$ defined by Eqs.~(\ref{dp-MSM-IFM})
and (\ref{L1-MSM-IFM}), respectively.

\begin{figure*}[t]
\includegraphics[width=\textwidth]{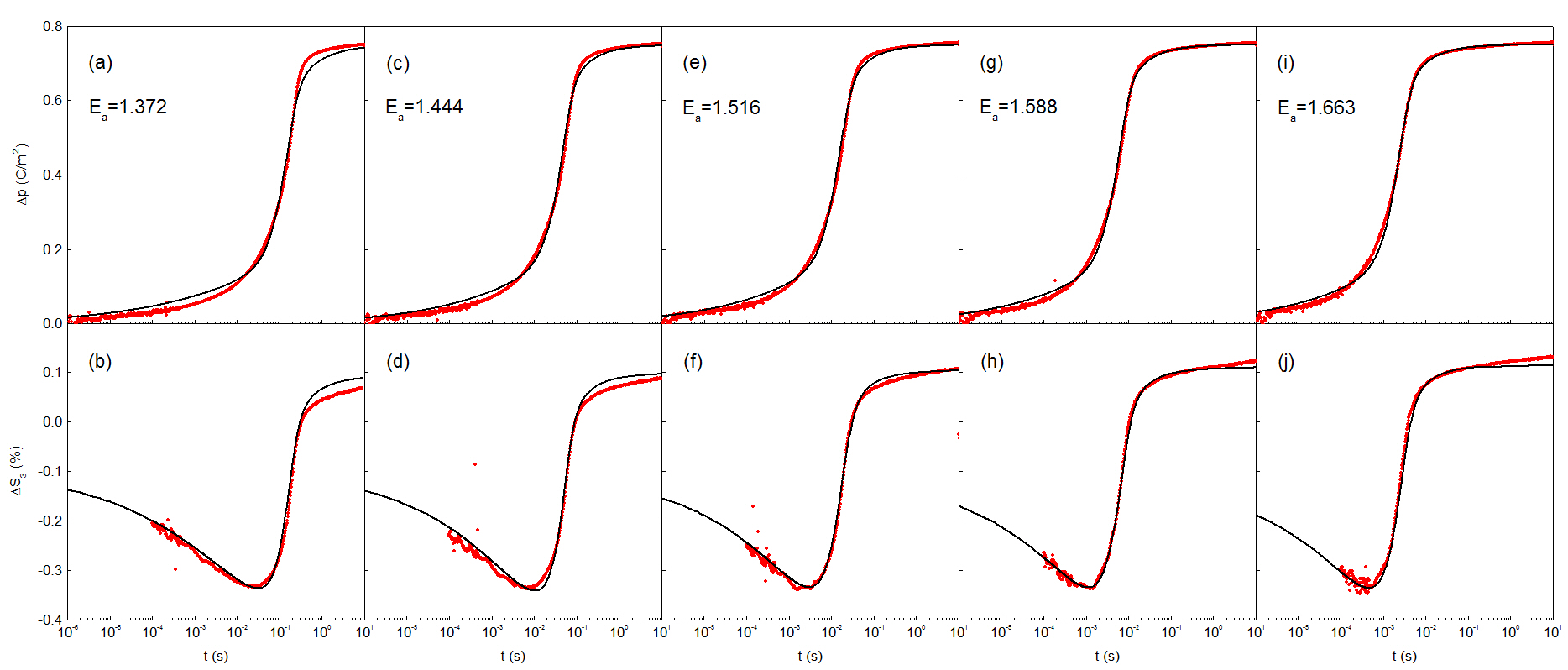}
\caption{Simultaneous variation of the polarization (a), (c), (e), (g), (i) and strain 
(b), (d), (f), (h), (j) with time at different field values in kV/mm  as indicated in
the plots. Experimental curves are shown by red symbols and fitting by means of the MSM-NLS 
model by black solid lines.}
\label{P-S-NLS}
\end{figure*}

The formulas (\ref{dS-MSM}-\ref{L1-MSM-IFM}), from now on termed as the MSM-NLS approach,
were used to fit simultaneous temporal measurements of polarization and strain response of 
\linebreak\noindent
Pb$_{0.985}$La$_{0.01}$(Zr$_{0.475}$Ti$_{0.525}$)O$_3$ ceramic with tetragonal phase 
symmetry\cite{GenenkoPRB2018-2}. We note that the only additional fitting parameter 
in comparison with Eqs.~(\ref{dp-MSM}) and (\ref{L1-MSM}) of the MSM 
model was the field distribution width $\kappa$. The fitting results shown in Fig.~\ref{P-S-NLS} 
by solid lines for different applied fields are in much better agreement with experimental data, 
shown by symbols, than the previous calculations using the MSM model\cite{GenenkoPRB2018-2}
neglecting the dispersive features of the response.  
The fitting parameters used in all shown graphs are $P_s=0.38$ C/m$^2$, $\Delta S_{max}=-1\%$,  
$\varepsilon_{33}=2.85\times 10^3$,
$Q_{11}=0.038$ m$^4$/C$^2$, 
$\tau_0=0.8\times 10^{-11}$ s, $\alpha=0.28,\,\gamma=2$, 
$\beta = 3$ for $E_a < 1.5$ kV/mm and 2 for $E_a > 1.5$ kV/mm, 
$\eta=0.42$, $\kappa =0.012$, $E_A^{(1)}=29$ kV/mm, $E_A^{(2)}=32.6$ kV/mm, 
$E_A^{(3)}=32.5$ kV/mm. Macroscopic parameters are in reasonable agreement with independently 
measured polarization and strain characteristics\cite{GenenkoPRB2018-2}. The determination of 
the microscopic fitting parameters has an inaccuracy below 1.3\% for $\alpha,\beta,\gamma,\kappa$ 
and $\eta$, and below 0.5\% for activation fields $E_A^{(i)}$. Note that the best fitting of 
experimental data was reached for the practically coinciding model parameters $E_A^{(2)}$ and 
$E_A^{(3)}$, followed by the coinciding characteristic times of the model $\tau_2$ and $\tau_3$.

\begin{figure*}[t]
\includegraphics[width=\textwidth]{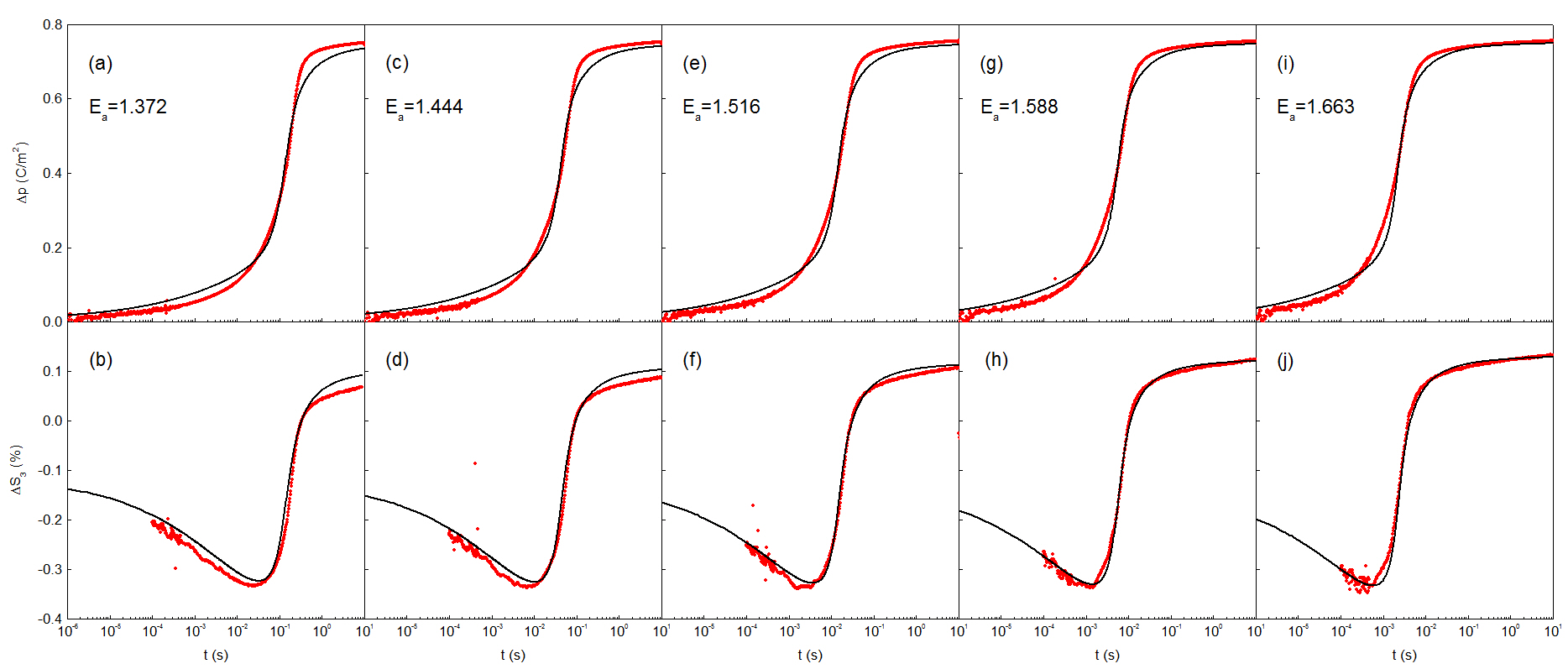}
\caption{Simultaneous variation of the polarization (a), (c), (e), (g), (i) and strain 
(b), (d), (f), (h), (j) with time at different field values in kV/mm  as indicated in
the plots. Experimental curves are shown by red symbols and fitting by means of the MSM-IFM 
model by black solid lines.}
\label{P-S-IFM}
\end{figure*}

In spite of the satisfactory description of the experiment by the formulas 
(\ref{dS-MSM}-\ref{L1-MSM-IFM}) they contain triple integration and are rather cumbersome for 
data fitting. 
In a more simple approach, conceptually closer to the previous IFM model\cite{IFM1,IFM2}, 
the local switching processes characterized by Avrami indexes larger than $2$ can be substituted
by step-functions on the logarithmic time scale\cite{IFM2} allowing for explicit integration 
over the local field variables in Eqs.~(\ref{dS-MSM}-\ref{L1-MSM-IFM}). This results in a simpler 
form, from now on termed as the MSM-IFM model,
\begin{align}
\label{dp-MSM-IFM-short}
\Delta p(t) &= 2P_s\eta \left\{1-\exp{\left[-\left(\frac{t}{\tau_1}\right)^{\alpha}\right]}\right\}
\nonumber \\
&+ 2P_s(1-\eta) \left\{\frac{1}{2} + \frac{1}{\pi}\arctan{\left[ \ln{\left(\frac{t}{t_3}\right)}/W_3\right]}
\right\}\nonumber\\
&-P_s\eta\frac{\alpha }{\tau_1}\int\limits_0^t dt_1  \left(\frac{t_1}{\tau_1}\right)^{\alpha -1}
\exp{\left[- \left(\frac{t_1}{\tau_1}\right)^{\alpha}\right]}\nonumber\\
&\times \left\{
\frac{1}{2} - \frac{1}{\pi}\arctan{\left[ \ln{\left(\frac{t-t_1}{t_2}\right)}/W_2\right]} \right\}.
\end{align}
containing only one integration over the intermediate moment $t_1$, which is unavoidable
when describing the convolution of two sequential switching events. Here, the ansatz
$\tau_1 = \tau _0 \exp{\left(E_A^{(1)}/E_a\right)}$ is retained while the characteristic times
$t_2$ and $t_3$ are assumed equal to each other together with their activation fields, 
\begin{equation}
\label{chartime}
t_{2,3} = \tau_0 \exp{\left(  \frac{E_A^{(2)}}{E_a (1+\kappa^2 )} \right)}\phantom{0}\text{and}\phantom{0}
W_{2,3} = \frac{\kappa}{1+\kappa^2} \frac{E_A^{(2)}}{E_a},
\end{equation}
according to the above fitting of the experimental data by the MSM-NLS model. The resulting description 
of the experiment shown in Fig.~\ref{P-S-IFM} appears to be of inferior quality to that of the MSM-NLS 
model, shown in Fig.~\ref{P-S-NLS}, however, the simpler MSM-IFM model also well describes the main 
features of polarization and strain kinetics, especially at high fields. Several parameters of the 
model thereby get somewhat modified, namely, $E_A^{(2)}=E_A^{(3)}=32.4$ kV/mm, 
$\eta =0.40$, $Q_{11}=0.044$ m$^4$/C$^2$, $\kappa =0.023$.
The fair quality of the fitting may result from the relatively narrow field distribution function,
Eq.~(\ref{Lorentz}), which limits the validity of the used step-function
approximation\cite{GenenkoPRB2018-2}. It is expected that the performance of the MSM-IFM approach will
be better for systems with broader field distributions.

Introduction of the statistical distribution of local field values remarkably improved the 
description of the electromechanical response, as compared with the MSM model\cite{GenenkoPRB2018-2}  
assuming the uniform electric field. The dispersion of the field values and, consequently, of the
local switching times, has the highest impact on the later stage of switching. However, particularly
this switching stage with a quasilinear behavior on the logarithmic time scale in Figs.~\ref{P-S-NLS} 
and \ref{P-S-IFM} is still not always properly captured by the both MSM-NLS and MSM-IFM models,
especially at low applied fields. It was suggested that this stage may appear due to a long-time 
electromechanical creep of the domain structure\cite{LiuJECS2006,SchultheissAM2018}. It is known, 
on the other hand, that such 
time dependences typically occur due to asymmetric field distributions enhanced in the low field 
region\cite{IFM1,ZhukovAPL2013,ZhukovJAP2015,ZhukovJAP2016,ZhukovAPL2016}, the feature missing here 
because of the simplifying choice of the symmetric Lorentzian distribution, Eq.~(\ref{Lorentz}). Using 
more realistic asymmetric field distributions may improve the performance of both models in the 
low field region.

Concerning the switching times governed by the respective activation fields, the considerably
shorter switching times $\tau_1(E_a)$ and the smallest activation field $E_A^{(1)}$ of the first 
90$^{\circ}$-switching processes are probably due to mechanical support by release of residual 
stresses, as suggested by Daniels {\it et al.}\cite{Daniels2014}. Furthermore, we note an astounding 
fact of the virtually identical activation fields $E_A^{(2)}$ and $E_A^{(3)}$ attributed to the 
presumably physically different -- 90$^{\circ}$ and 180$^{\circ}$ -- switching processes which are 
characterized by rather different activation energies\cite{LiuNature2016}. This coincidence 
may be related to the hypothesis by Arlt\cite{Arlt1997}, who suggested a possible scenario of the 
coherent 90$^{\circ}$ switching events, which do not contribute to the spontaneous strain, but 
contribute to the polarization and thus may be experimentally mistaken for 180$^{\circ}$ processes. 
For this to occur, these processes should be correlated over mesoscopic length scales depending on 
the microstructural properties of materials. We note that the formulas 
(\ref{dp-MSM},\ref{L1-MSM},\ref{dp-MSM-IFM},\ref{L1-MSM-IFM},\ref{dp-MSM-IFM-short}) do not assume
explicitly, but also do not preclude, correlated switching processes. Generally, 
atomistic\cite{LiuNature2016} simulations in uniform media are strongly in favor of highly correlated 
90$^{\circ}$ switching processes, coherent over macroscopic scales, which is confirmed by direct 
optical observations on a BaTiO$_3$ single crystal\cite{JiangAPL2008}. Macrosopically coherent 
switching is also predicted by other atomistic\cite{Leschhorn2017} and phase-field\cite{ZhouJAP2012} 
simulations for single crystals. The study of correlations 
in the polarization response of thin ferroelectric films revealed coherent behavior of up to 1000
grains\cite{BintachittAPL2009,SealPRL2009} which, however, may be mediated by elastic coupling
through the substrate\cite{BintachittPNAS2010,GriggioPRL2012}. Switching processes in bulk ceramics
seem to be rather correlated at a short-range scale involving around 20 
grains\cite{DanielsSciRep2016,MajkutJACS2017} which roughly corresponds to the number of the next
neighbors. Similar conclusions on the correlation length scale were derived 
from 2D\cite{KhachaturyanPRB2017} and 3D\cite{KhachaturyanPRB2018} simulations where,
however, only electric interactions where taken into account. Polarization correlations of 
neighboring grains due to domains crossing grain boundaries were observed by optical 
observations\cite{Devries1957}, TEM and PFM\cite{MarincelAFM2014}.

In conclusion, by supplementing the recent multistep stochastic mechanism (MSM) model of 
polarization switching in ferroelectrics\cite{GenenkoPRB2018-2} with the statistical 
distribution of local electric fields,
the new hybrid MSM-NLS model (with its simplified MSM-IFM version) was advanced. 
The new model allows description of the simultaneous polarization and strain response of
ferroelectric ceramics over a wide time domain with high accuracy that was exemplarily shown
for a tetragonal PZT ceramic. Particularly, it allows determination of the fractions of 
sequential 90$^{\circ}$- and parallel 180$^{\circ}$-switching events $\eta$. However, the analysis 
of the model parameters resulting from the fitting of experiments revealed a notable fact 
that the switching time and activation field for the second sequential 90$^{\circ}$-switching 
processes coincide with those of the parallel 180$^{\circ}$-switching events. 
The contribution of the latter into polarization and strain can hardly be distinguished 
from such coherent 90$^{\circ}$-switching processes, which do not contribute to the spontaneous 
strain\cite{Arlt1997}. Though the present model, as well as other common statistical models, 
are based on the assumption of statistically-independent switching events, the coincidence 
of the switching times and activation fields of the sequential 90$^{\circ}$- and parallel 
180$^{\circ}$-switching processes can hardly be accidental. This might indicate that the 
polarization reversal is rather dominated by a mix of statistically independent and coherent 
90$^{\circ}$-switching events, correlated on different length-scales within and beyond the 
grains, than by parallel 180$^{\circ}$-switching events.\\

This work was supported by the Deutsche Forschungsgemeinschaft (DFG) Grants No. GE 1171/7-1 
and No. KO 5100/1-1.

\vspace{5mm}
\bibliographystyle{plain}
\bibliography{apssamp}

\end{document}